\documentclass[draft]{agujournal2019}
\usepackage{url} 
\usepackage{lineno}

\usepackage{todonotes}
\usepackage{soul}
\usepackage[english]{babel}
\usepackage{amsmath,amssymb,bm}
\usepackage{float}
\usepackage[section]{placeins}

\newcommand{\cla}{\mathtt{A}} 
\newcommand{\clb}{\mathtt{B}} 
\newcommand{\clc}{\mathtt{C}} 
\newcommand{\claa}{\mathtt{AA}}
\newcommand{\clab}{\mathtt{AB}}
\newcommand{\clbb}{\mathtt{BB}}
\newcommand{\clba}{\mathtt{BA}}

\makeatletter
\renewcommand\subsubsection{%
    \@startsection{paragraph}{4}{\z@}%
        {0pt}
        {-1em}
        {\normalfont\normalsize\bfseries}%
}%
\makeatother

\draftfalse

\journalname{JGR: Machine Learning and Computation}

\begin{document}

\title{AI Emulation of Stochastic Sudden Stratospheric Warming with Interpretable Latent Structure}

%
%




\authors{
C.\ Daniel Boscu\affil{1}$^{*}$,
Daniel Hernandez\affil{1}$^{*}$,
Fabio Alvarez Ventura\affil{1}$^{*}$,
Justin Finkel\affil{1,2},
Ashesh Chattopadhyay\affil{3},
Pedram Hassanzadeh\affil{1,2,4},
and Dorian S.\ Abbot\affil{1}
}

\begin{center}
$^{*}$These authors contributed equally to this work.
\end{center}


\affiliation{1}{Department of Geophysical Sciences, University of Chicago, Chicago, 60637, IL}
\affiliation{2}{Data Science Institute, University of Chicago, Chicago, 60637, IL}
\affiliation{3}{Department of Applied Mathematics, University of California, Santa Cruz, Santa Cruz, 95064, CA}
\affiliation{4}{Committee on Computational and Applied Mathematics, University of Chicago, Chicago, 60637, IL}




\correspondingauthor{Dorian S. Abbot}{abbot@uchicago.edu}



\begin{keypoints}
\setlength{\itemsep}{0.5em}
\item We probabilistically emulate the stochastic Holton-Mass model of sudden stratospheric warming events with a CVAE.
\item The emulator reproduces key short and long-term sudden warming statistics, including regime persistence, committor, and expected lead times.
\item Latent space analysis reveals four well-separated, physically interpretable dynamical regimes.
\end{keypoints}

%
%

%
%


\begin{abstract}
Rare weather regime transitions pose a challenge for data-driven modeling due to class imbalance. In this study, we develop a probabilistic deep learning emulator for a prototypical system with regime transitions, the stochastic Holton--Mass model of stratospheric variability, and analyze the structure of its learned latent space. The Holton--Mass model exhibits two metastable regimes, a strong and a weak polar vortex, maintained by nonlinear wave--mean flow interactions, with weak stochastic forcing intermittently triggering rare transitions between these regimes that qualitatively represent SSW events. We employ a ResNet-inspired Conditional Variational Autoencoder with six-layer encoder and decoder layers and explicit current-state conditioning to model the distribution of the system's state at the next time step (one day). The emulator accurately reproduces short-term dynamics, steady-state probability distributions, regime persistence statistics, rare transition rates, the transition committor function, and the transition expected lead time of the physical model. Beyond emulation fidelity, we interrogate the learned latent representation to understand how the model internalizes the underlying metastable structure of the dynamics. Principal Component Analysis of the 32-dimensional latent space reveals a clear and unsupervised separation into four physically interpretable clusters corresponding to strong versus weak vortex regimes and stable versus transition-prone configurations. Such emergent regime separation in latent space is hard to identify for deep generative models applied to high-dimensional stochastic systems. Our results show that carefully designed probabilistic emulators can uncover physically meaningful manifolds governing extreme-event dynamics, potentially aiding the development of improved operational advanced warning systems.
\end{abstract}


\section*{Plain Language Summary}
Sudden Stratospheric Warmings (SSWs) are disruptions of the polar vortex, which are intermittent, difficult to predict, and lead to extreme winter cold across Europe and North America. Here we train an Artificial Intelligence emulator on the Holton-Mass model, an idealized stratospheric model that qualitatively reproduces SSWs. This model is stochastic, or randomly forced, in contrast with most atmospheric models that emulators have been built for. The model has two metastable, or quasi-stable, regimes that it spends long times in, until the random forcing happens to drive a transition, corresponding to an SSW. Although our emulator is trained to advance the state of the system by one day, it captures long-time statistics of the model including transition statistics and detailed physical transition pathways. Crucially, we are able to identify a clean, interpretable signal in the interior of the emulator that predicts whether a transition is imminent or not, which is a rare and exciting result for an Artificial Intelligence emulator. This suggests a methodology for building an operationally useful diagnostic for SSWs and other types of transitions.


\section{Introduction}
\label{sec:introduction}

The stratospheric polar vortex is a key component of polar and midlatitude winter climate, particularly on subseasonal-to-seasonal timescales \cite{wcd-5-231-2024}. Its dynamical importance is most evident during sudden stratospheric warming (SSW) events, in which a rapid breakdown of the polar vortex reorganizes the large-scale circulation, alters the tropospheric jet stream \cite{BaldwinDunkerton2001}, and can trigger persistent extreme cold outbreaks at the surface \cite{https://doi.org/10.1002/2015JD023860, Kautz2020enhanced}. Accurately forecasting these events remains challenging because of the need for good representation of coupling between the stratosphere and troposphere and good resolution throughout the stratosphere, to represent the diverse range of driving processes operating at different timescales from days (e.g., convection) to years (e.g., the quasi-biennial oscillation and El Ni{\~n}o southern oscillations), which many models have developed only recently \cite{Domeisen2020role, Chwat2022which,Garfinkel2025process}. SSWs are an example of \emph{weather regime transitions} \cite{Liang2023northern}, other examples including blocking and persistent jet configurations. Weather regimes generally arise from multiscale coupled processes, and transitions are challenging to predict because of their abruptness and relative sparsity in the historical record compared to everyday phenomena, like midlatitude cyclones. 

The stochastic Holton--Mass (HM) model \cite{Holton1976stratospheric} is an early and influential discretized physical model of the polar vortex (Fig.~\ref{fig:all-in-one}a gives a visual summary of its main dynamical behavior). Despite its idealized nature, the HM model captures essential features of stratospheric variability through nonlinear wave--mean flow interactions and provides a physically interpretable framework for studying vortex dynamics and SSWs. In the HM model the state of the system at time $t$ is represented by a 75-dimensional vector $\mathbf{X}(t)$ that includes a complex-valued perturbation streamfunction $\Psi(z)$ and the zonal-mean zonal wind $U(z)$, discretized over 25 vertical levels with uniform spacing in the log-pressure coordinate $z$:
\begin{equation}
\label{eq:state_vector}
\begin{aligned}
\mathbf{X}(t) = [\,&\mathrm{Re}\{\Psi(\Delta z, t)\}, \ldots, \mathrm{Re}\{\Psi(z_{\mathrm{top}} - \Delta z, t)\}, \\
&\mathrm{Im}\{\Psi(\Delta z, t)\}, \ldots, \mathrm{Im}\{\Psi(z_{\mathrm{top}} - \Delta z, t)\}, \\
&U(\Delta z, t), \ldots, U(z_{\mathrm{top}} - \Delta z, t)\,] \\
= [&\mathbf{\Psi}(t),\mathbf{U}(t)]\\
\in &\,\mathbb{R}^d = \mathbb{R}^{75},
\end{aligned}
\end{equation}
where bold-face denotes a vector quantity. 
Detailed descriptions of the governing equations, truncation, and physical interpretation can be found in \citeA{Holton1976stratospheric,Yoden1987dynamical,Finkel21,Finkel2022data}.

In the absence of stochastic forcing, the HM system admits two stable equilibria corresponding to a strong and a weak polar vortex. 
These equilibria arise from nonlinear wave--mean flow interactions coupled with height-dependent radiative cooling \cite{Holton1976stratospheric}. The strong-vortex equilibrium has large positive (eastward) zonal wind background $U$ that increases steadily with altitude, and substantial but stable waves, $\Psi$. The weak-vortex equilibrium has greatly diminished zonal wind $U$ that nearly vanishes at $z\approx30$ km, and waves, $\Psi$, with a large phase tilt across altitudes, which can drive vertical transport of westward momentum and cancel out the stabilizing force of radiative cooling. This environment is conducive to repeated ``vacillation cycles'' \cite{Holton1976stratospheric} where the vortex repeatedly partially recovers and then gets undercut. 

When stochastic forcing is added to represent fast, unresolved processes such as gravity-wave drag, the system becomes metastable: trajectories spend long intervals fluctuating near one equilibrium before rare, intermittent transitions to the other (Fig.~\ref{fig:timeseries_pdf}). Transitions of the vortex from strong to weak qualitatively represent SSW events \cite{Birner2008sudden,Finkel21}.

Recent advances in machine learning (ML) emulators have demonstrated substantial promise for weather \cite{pathak2022fourcastnet,bi2023accurate,lam2023learning}
and SSW \cite{du2025predicting,Tao2026efficient} forecasting. Emulators trained on models have focused on deterministic (but chaotic) systems. The emulation of stochastic models of atmospheric dynamics, however, has been investigated less thoroughly. In this work we set out to accomplish two tasks: (1) build an emulator for a meta-stable system with noise-driven transitions and (2) demonstrate a physically interpretable latent space of this emulator. We use the HM model as a controlled testbed with clear physical interpretation and well-characterized regime behavior, and establish that critical physical processes are being encoded correctly by checking essential diagnostics that have been previously identified for the HM model.

Accurately modeling such metastable dynamics requires a stochastic ML framework. A deterministic emulator would underestimate variability and collapse onto a single regime conditioned on the initial state. We therefore employ a Conditional Variational Autoencoder (CVAE) to sample from the conditional distribution of plausible future states, $P(\mathbf{X}(t+\Delta t)\,|\,\mathbf{X} (t))$, rather than producing a deterministic forecast \cite{Chattopadhyay23, sambamurthy2025lazy,wyrod2025generative}.  Importantly, we do not take advantage of our knowledge of the noise structure in the HM model; instead, we allow the CVAE to learn it.

Rare transitions in a metastable system represent an archetypal data imbalance problem for data-based modeling \cite{krawczyk2016learning}: almost all of the training data is in one of the metastable sets, and very little is from a rare transition between them. AI weather models have shown mixed success on the rarest extremes, at times able to extrapolate locally using information from other spatial locations, but typically struggling to extrapolate to situations entirely outside of the training data \cite{hakim2024dynamical,sun2025predicting,zhang2025numerical,sun2025can}. Given that transitions are present, albeit rare, in the training data of our target application, this previous work suggests a CVAE may be able to learn the HM model dynamics.

Beyond reproducing observable statistics, we aim to understand what the emulator internalizes about the system dynamics. We therefore analyze the structure of the learned latent space, similar to~\citeA{shamekh2023implicit}. Similar analysis has been done across other studies in dynamical systems and fluid mechanics, e.g., in~\citeA{magri2022interpretability} and~\citeA{page2021revealing}. Using principal component analysis (PCA), we demonstrate that distinct dynamical regimes—strong vortex, weak vortex, and transition-prone states—naturally separate into clusters without explicit supervision. This emergent organization provides insight into how the emulator represents metastability and transition pathways and offers a diagnostic tool for interpreting deep generative models applied to stochastic climate dynamics that could potentially be developed into an operational early warning system for rare regime transitions.

\vspace{-1em}

\section{Methods}

\begin{figure}[h!]
    \vspace{-1em}
    \centering
    \includegraphics[width=1\linewidth]{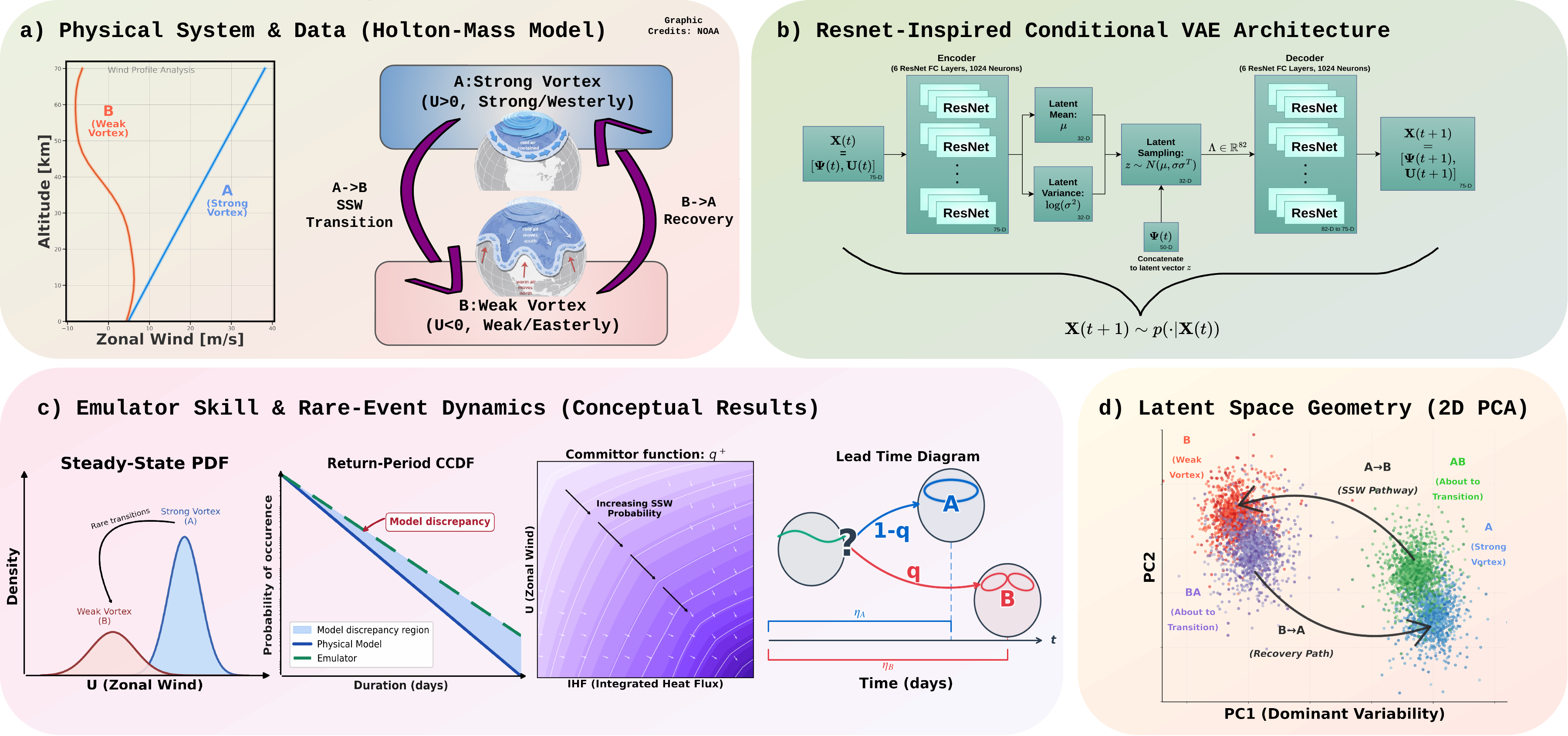}
    \caption{The research pipeline. (a) Visual representation of the Holton-Mass model data and the metastable system of the Polar Vortex. (b) Architecture of the  Conditional Variational Autoencoder (CVAE) emulator.
    The encoder $q_{\boldsymbol{\phi}}(z\,|\,\mathbf{X}(t))$, implemented as a six-layer residual fully connected network with learnable parameters $\boldsymbol{\phi}$,
    maps the current system state $\mathbf{X}(t)$ to a 32-dimensional latent mean and log-variance.
    A latent variable $z$ is sampled and concatenated with the resolved streamfunction perturbation field $\boldsymbol{\Psi}(t)$.
    The decoder $p_{\boldsymbol{\theta}}(\mathbf{X}(t+\Delta t)\,|\,\boldsymbol{\Psi}(t),z)$,
    also implemented as a six-layer residual fully connected network,
    outputs the predicted next state at one-day expected lead time.
    This probabilistic architecture enables the emulator to represent the metastability, stochastic variability,
    and noise-induced regime transitions of the Holton-Mass model. (c) Four techniques we use to compare the long-term behavior of the Emulator with the Holton-Mass model: the state PDF, the return period complementary cumulative distribution function ($\text{CCDF}(\tau)=1-\text{CDF}(\tau)=\mathbb{P}\{\text{return period}>\tau$), the transition probability (committor function), and the transition lead time. (d) PCA analysis of the emulator's latent mean vector, $\mu$, which clusters into the four dynamical categories.}
    \label{fig:all-in-one}
\end{figure}

This section describes the pipeline used in this study as shown in Figure~\ref{fig:all-in-one}, including the construction of the training dataset from the stochastic Holton--Mass model, the architecture of the probabilistic emulator, and the training and model-selection procedures used to ensure both short-term predictive skill and long-term statistical fidelity.

\subsection{Data}
\label{sec:data}

The training dataset consists of a $3\times10^{5}$-day simulation of the stochastic Holton--Mass model as implemented by \citeA{Finkel21} and specified in their Eqs. 5-8. The most relevant integration parameters for our emulation pipeline are the timestep ($0.005$ days, with the Euler-Maruyama scheme) and the sampling frequency of once per day. The many timesteps within a single sample renders the learning target ($\mathbf{X}(t)\mapsto\mathbf{X}(t+1)$) a more complicated distribution than a simple Gaussian, making for a more challenging learning problem.

Following prior work \cite{Finkel21}, we define the strong and weak vortex regimes as extended sets within state space, denoted $A$ and $B$, using thresholds on the zonal wind at 30 km:
\begin{align}
A &= \{\mathbf{X}=[\mathbf{\Psi},\mathbf{U}] : U(30\,\mathrm{km}) \ge u_A := 53.8~\mathrm{m}\,\mathrm{s}^{-1}\}, \label{eq:basin_A} \\
B &= \{\mathbf{X}=[\mathbf{\Psi},\mathbf{U}] : U(30\,\mathrm{km}) \le u_B := 21.4~\mathrm{m}\,\mathrm{s}^{-1}\}. \label{eq:basin_B}
\end{align}
The thresholds $u_A$ and $u_B$ are chosen based on the locations of the stable equilibria of the unforced system, sometimes referred to as equilibrium points $\mathbf{a}$ and $\mathbf{b}$. Our results are not sensitive to moderate variations in these thresholds.

For later convenience, we introduce a parallel classification of days $\{t\}$ from a timeseries $\{\mathbf{X}(t)\}$, which applies to integrations of either the Holton--Mass model or the emulator: $t\in\cla$ when $\mathbf{X}(t)\in A$, $t\in\clb$ when $\mathbf{X}(t)\in B$, and $t\in\clc$ when $\mathbf{X}(t)\in C=(A\cup B)^c$. But because of the randomness in a particular timeseries, we further sub-categorize $\cla$ as either \emph{transient} or \emph{persistent}. At transient times $t\in\clab$, $\mathbf{X}(t)$ is about to leave $A$ bound for $B$ in the very next timestep. At persistent times $t\in\claa$, $\mathbf{X}(t)$ either stays within $A$ at the next timestep or is about to leave $A$ and then come back without visiting $B$. The transient class $\clba$ and persistent class $\clbb$ are defined analogously. 

The transient class is defined to isolate the start of complete regime changes rather than brief excursions. To be more precise, $t\in\clab$ if the following conditions hold:
\begin{enumerate}
    \item $\mathbf{X}(t)\in A$,
    \item $\mathbf{X}(t+1)\notin A$,
    \item the system reaches set $B$ before returning to set $A$.
\end{enumerate}
Formally, if we define the first hitting times for the two metastable sets as
\[
\tau_{\cla}(t) = \min\{s \ge t : \mathbf{X}(s)\in A\}, \qquad
\tau_{\clb}(t) = \min\{s \ge t : \mathbf{X}(s)\in B\},
\]
the third condition of the $\clab$ class can be stated $\tau_{\clb}(t+1) < \tau_{\cla}(t+1)$. The $\clba$ class is defined analogously. These criteria ensure that $\clab$ and $\clba$ samples correspond to genuine regime transitions rather than ephemeral fluctuations back and forth across thresholds that don't lead to a shift between metastable sets. 

Figure~\ref{fig:timeseries_pdf} shows representative timeseries and probability density functions of the zonal velocity at an altitude of 30 km, $U(30\,\mathrm{km})$, for the Holton--Mass model and the emulator.

\subsection{Probabilistic Emulator Architecture}
\label{sec:architecture}

The Holton--Mass system is stochastic and metastable, characterized by long-lived dynamical regimes separated by noise-induced transitions. Consequently, a useful emulator must represent a \emph{distribution} over future states rather than a single deterministic forecast. We therefore model the conditional distribution of the next system state using a Conditional Variational Autoencoder (CVAE).

The CVAE consists of two components. The encoder, parameterized by weights $\boldsymbol{\phi}$, defines an approximate posterior distribution
\[
q_{\boldsymbol{\phi}}\!\left(z \,\middle|\, \mathbf{X}(t)\right),
\]
where the latent variable $z\in\mathbb{R}^{32}$ represents unresolved stochastic variability. The decoder, parameterized by weights $\boldsymbol{\theta}$, defines the conditional distribution
\[
p_{\boldsymbol{\theta}}\!\left(\mathbf{X}(t+\Delta t) \,\middle|\, \boldsymbol{\Psi}(t), z\right),
\]
where $\Delta t=1$ day is the output interval.

Conditioning is implemented by concatenating the sampled latent variable $z$ with the resolved perturbation streamfunction, $\boldsymbol{\Psi}(t)$, before it enters the decoder. This design re-injects physically relevant context directly into the generative pathway, ``reminding'' the decoder of where the latent variable, $z$, came from originally before ``forgetting'' initial-state information to represent unresolved stochastic influences. We do not include the zonal wind $\mathbf{U}(t)$ in the conditioning input, as including it degraded long-term statistical fidelity in exploratory experiments.

Both the encoder and decoder are implemented as residual multilayer perceptrons (ResNet--MLPs). Each network consists of six fully connected layers of width 1024 with ReLU activations and identity skip connections. For either network, the hidden layers compute
\[
\mathbf{h}_{\ell+1}
=
\mathrm{ReLU}\!\left(W_\ell \mathbf{h}_\ell + \mathbf{b}_\ell\right)
+
\mathbf{h}_\ell,
\]
which preserves gradient flow and enables stable learning of both near-identity and strongly nonlinear mappings.

The encoder outputs a latent mean $\boldsymbol{\mu}(\mathbf{X}(t))\in\mathbb{R}^{32}$ and log-variance $\log\boldsymbol{\sigma}^2(\mathbf{X}(t))\in\mathbb{R}^{32}$, defining a diagonal Gaussian posterior. A latent sample is obtained via the reparameterization
\[
z = \boldsymbol{\mu} + \boldsymbol{\sigma} \odot \boldsymbol{\varepsilon},
\qquad
\boldsymbol{\varepsilon}\sim\mathcal{N}(\mathbf{0}, I_{32}).
\]
The decoder maps the concatenated vector $[z;\boldsymbol{\Psi}(t)]\in\mathbb{R}^{82}$ to a predictive sample of the next state $\mathbf{X}(t+\Delta t)$.

\vspace{-0.5em}
\subsection{Training and Model Selection}
\label{sec:training}

Training minimizes the evidence lower bound (ELBO),
\[
\mathcal{L}(\boldsymbol{\theta},\boldsymbol{\phi})
=
\mathbb{E}_{q_{\boldsymbol{\phi}}(z\,|\,\mathbf{X}(t))}
\!\left[\log p_{\boldsymbol{\theta}}(\mathbf{X}(t+\Delta t)\,|\,\boldsymbol{\Psi}(t),z)\right]
-
\beta\,D_{\mathrm{KL}}\!\left(q_{\boldsymbol{\phi}}(z\,|\,\mathbf{X}(t))\,\|\,p(z)\right),
\]
where $p(z)=\mathcal{N}(z|\mathbf{0},I_{32})$. The coefficient $\beta$ is annealed cyclically in the range $[0.01,0.3]$ to mitigate posterior collapse.

The reconstruction term is implemented using a robust Huber loss \cite{Huber},
\[
\mathcal{L}_{\mathrm{rec}}
=
\sum_{j=1}^{75}
\mathrm{Huber}_\delta\!\big((\mathbf{X}(t+\Delta t))_j-(\hat{\mathbf{X}}(t+\Delta t))_j\big),
\]
which reduces sensitivity to large residuals associated with rare transitions. The KL divergence has the standard closed-form expression for diagonal Gaussian distributions.

The emulator is trained on 250{,}000 days of data using mini-batches of size 1024, with a validation set of 50{,}000 days. This dataset size is chosen to ensure adequate sampling of rare events given the typical return period of vortex breakdowns, including in total 388 SSWs. We also trained an emulator with only 10\% of this training data, including 33 SSWs, and achieved equivalent performance on the climatological metrics introduced below. Training is performed for 1000 epochs using a learning rate of $10^{-4}$.

Because minimizing short-term reconstruction loss does not guarantee accurate long-term statistics, model selection is based on long-term climatological metrics, including errors in return-period distributions and the KL divergence between physical and emulated steady-state probability density functions. Multiple training runs are performed to account for stochastic variability, and the epoch minimizing the aggregate long-term error is selected. All long-term statistics are computed using up to $10^6$ days of data, as the emulator became unstable beyond 1.4 million days due to runaway error propagation.

A final fine-tuning step corrects a slight over-representation of the strong-vortex regime in the emulator. After identifying excess samples in regime $A$, a subset is removed from the training data and the model is fine-tuned using a reduced learning rate of $10^{-8}$. This procedure reduces the KL divergence error from 0.119 to 0.097, corresponding to an approximately $18.5\%$ increase in captured information relative to the physical distribution.

\vspace{-1em}
\begin{figure}[h!]
    \centering
    \includegraphics[width=1\linewidth]{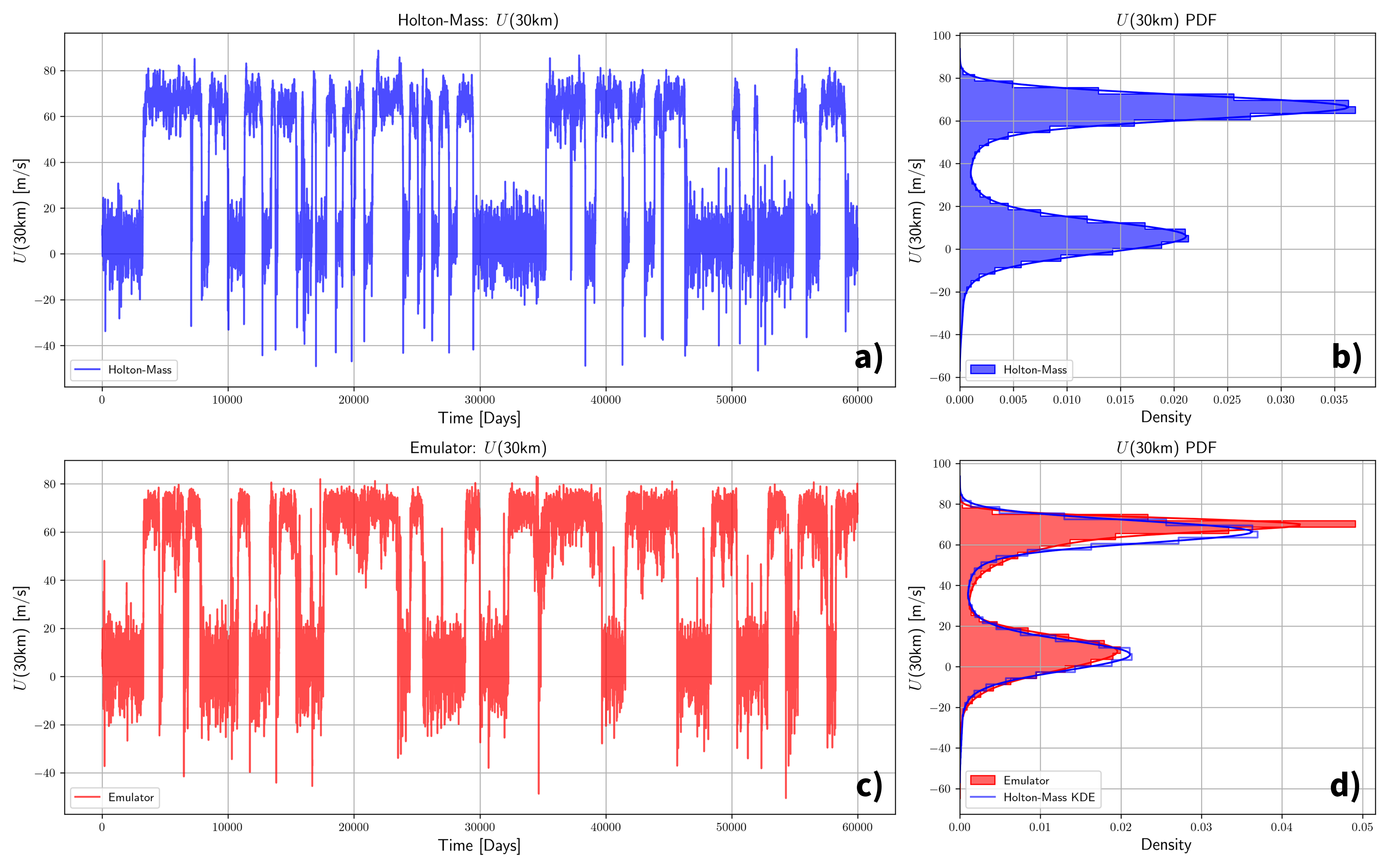}
    \caption{Sample timeseries (left) and Probability Density Function (right, PDF) of the zonal velocity at an altitude of 30 km, $U$(30 km), for the Holton-Mass Model (blue, $6 \times 10^4$ days) and the emulator (red, $10^6$ days). The PDF of the emulator is accompanied by the kernel density estimate (KDE) of the Holton-Mass model's PDF. }
    \label{fig:timeseries_pdf}
\end{figure}
\section{Results}

\subsection{Short-term Accuracy} 

We can quantify the forecast skill of the emulator using the RMSE:
\begin{align}
    \text{RMSE}_i(t) &= \sqrt{\frac{1}{N_s} \sum_{s=1}^{N_s} \left( \bar{y}_{i,t,s} - y_{i,t,s}^{\text{true}} \right)^2}\\
    \bar{y}_{i,t,s} &= \frac{1}{N_e} \sum_{e=1}^{N_e} y_{i,t,s,e}\\
    \overline{\text{RMSE}}(t) &= \frac{1}{N_{\text{IC}}} \sum_{i=1}^{N_{\text{IC}}} \text{RMSE}_i(t)
\end{align}

\noindent where:
\begin{itemize}
    \item $t$: forecast lead time [days], $t \in \{0, 1, \ldots, 400\}$
    \item $N_s = 25$: number of spatial grid points (altitude levels)
    \item $N_e = 50$: number of ensemble members
    \item $N_{\text{IC}} = 1000$: number of initial conditions
    \item $y_{i,t,s,e}$: prediction at time $t$ after initial condition $i$, point $s$, ensemble member $e$ \hfill $(1000\times401 \times 25 \times 50)$
    \item $\bar{y}_{i,t,s}$: ensemble mean prediction at time $t$ after initial condition $i$, point $s$ $(1000\times401 \times 25)$
    \item $y_{t,s}^{\text{true}}$: ground truth at time $t$ after initial condition $i$, point $s$ $(1000\times401 \times 25)$
    \item $\text{RMSE}_i(t)$: RMSE at time $t$ after initial condition $i$  $(1000\times401,)$
    \item $\overline{\text{RMSE}}(t)$: mean RMSE across all initial conditions $(401,)$
\end{itemize}

\begin{figure}[h!]
    \centering
    \includegraphics[width=1\linewidth]{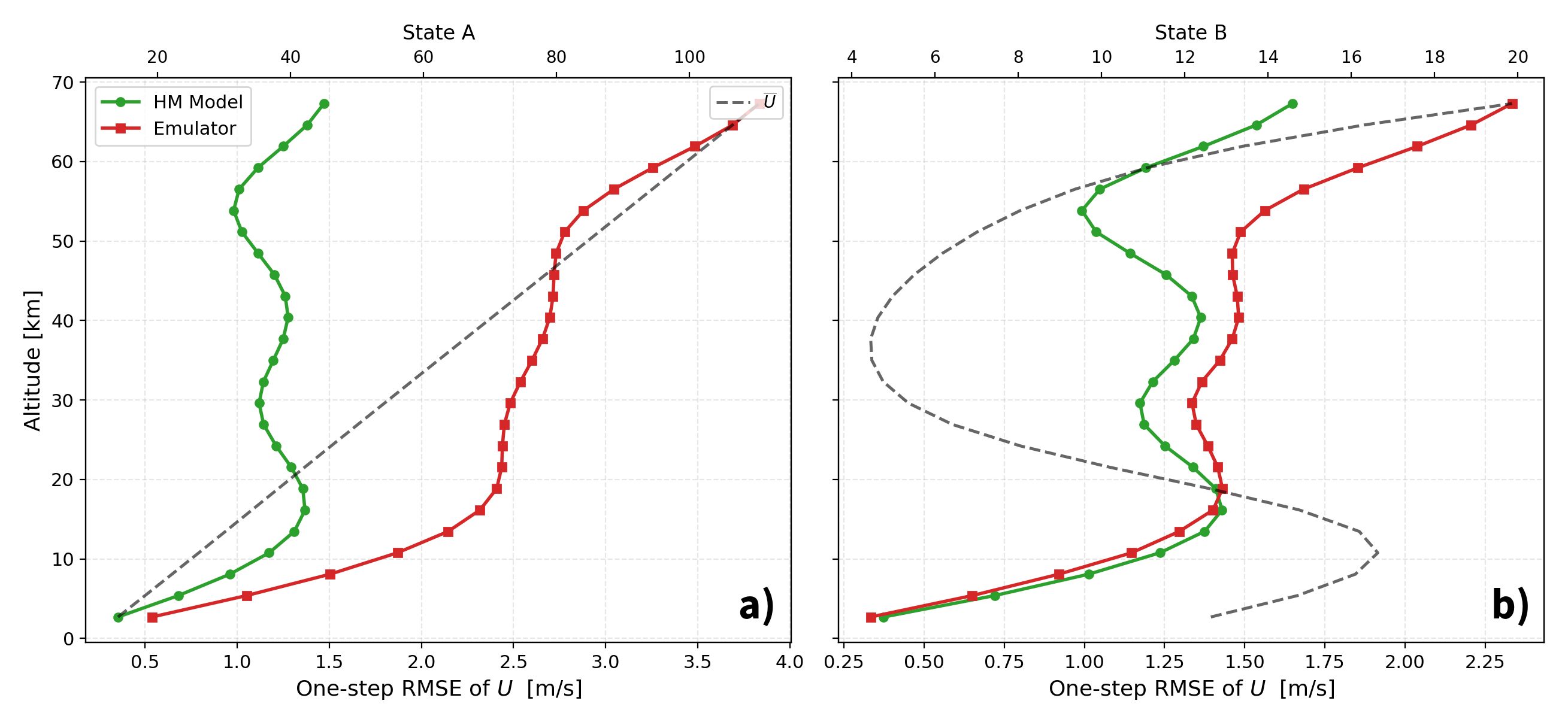}
    \caption{One-step RMSE of the zonal wind, $U$, as a function of altitude, $z$, computed from 500-member ensembles and averaged over 200 initial conditions sampled from each dynamical state. Results are shown for the emulator (red) and the Holton-Mass stochastic model (green) in state $A$ (left) and state $B$ (right). The dashed grey curves shows the climatological zonal wind $\bar{U}$ for each state.}
    \label{fig:one_step_updated}
\end{figure}

First we evaluate the shortest-term accuracy of the emulator by comparing the RMSE of the zonal velocity after one step for an ensemble of emulations with a similar ensemble of Holton-Mass model simulations (Fig.~\ref{fig:one_step_updated}). The RMSE of the Holton-Mass model gives us the one-step spread that a perfect emulator would have due simply to stochasticity. In state $B$ the one-step RMSE profile of the emulator closely tracks that of the Holton-Mass model across all altitudes. Both exhibit a non-monotonic increase in RMSE with altitude, with local maxima around 15 and 40 km.  In state $A$, the RMSE increases with altitude again, but the RMSE of the emulator is roughly twice as large as that of the Holton-Mass model at most altitudes. Overall we can conclude from this that while the one-step emulator performance is decent in both states, it is much better in state $B$ than state $A$. Moreover, stochastic forcing leads to more variance in zonal wind at higher than lower altitudes. 

\vspace{-1em}

\begin{figure}[h!]
\centering
\includegraphics[width=1\textwidth]{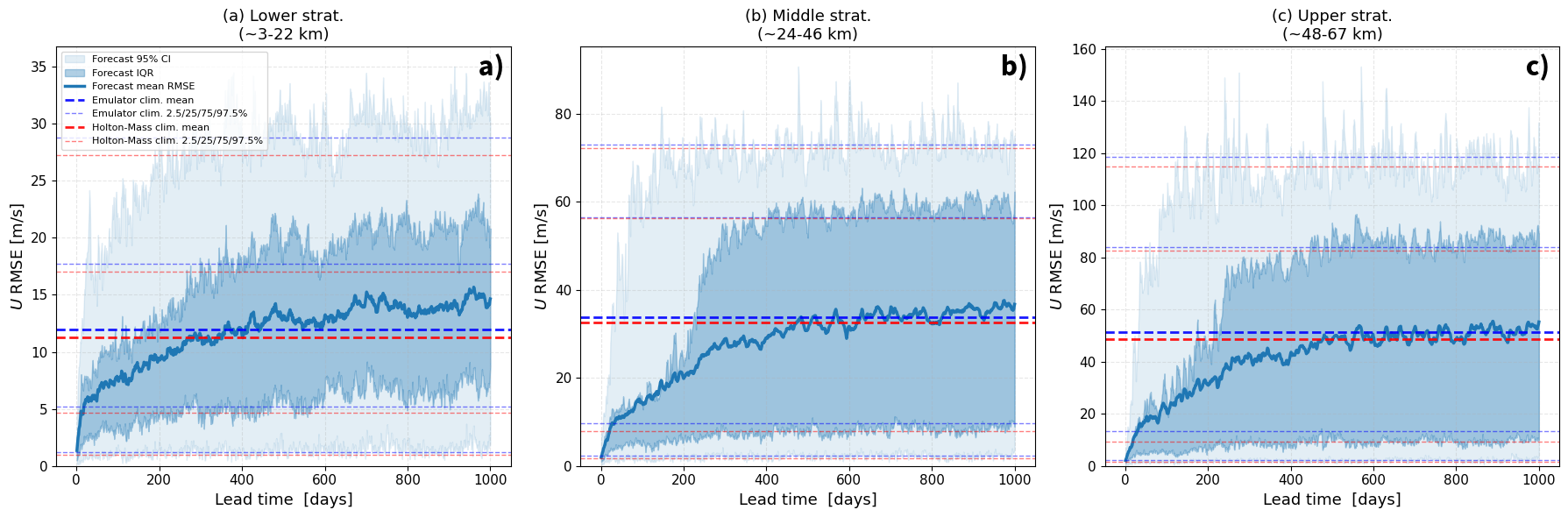}
\caption{Forecast error growth ($\overline{\text{RMSE}}$) averaged over 2000 initial conditions sampled from the model climatology. The dark blue line shows the mean  RMSE, with the medium blue region indicating the interquartile range (IQR, 25--75\%) and the light blue region showing the 95\% confidence interval. Horizontal dashed blue lines show the asymptotic values of each quantile. Red horizontal dashed lines show the corresponding quantiles for the Holton-Mass model, showing close agreement.  The sharp uptick in 75th-percentile error near 200 days for the mid-stratosphere is due to the onset of regime transitions at this timescale.}
\label{fig:forecast_error_growth}
\end{figure}

Next we consider the loss of emulator forecast skill with forecast time. The RMSE grows rapidly during the first $\sim$50 days as the ensemble forecasts diverge from the true trajectories, reflecting the chaotic nature of the system (Fig.~\ref{fig:forecast_error_growth}). Beyond approximately 1000 days, the error growth saturates. Most transitions have occurred on this timescale and specific, deterministic forecasts are no longer useful for this system. 

\subsection{Long-term Accuracy}

The PDFs in figure~\ref{fig:timeseries_pdf} represent one projection of the long-term dynamics, or ``climatology'' and demonstrate good agreement between the Emulator and the Holton-Mass model. The Emulator captures the distribution of zonal velocity at 30 km ($U(\text{30 km})$) in state $B$ very well and qualitatively reproduces the $U(\text{30 km})$ distribution in state $A$, with a slightly overestimated mean and a slightly underestimated spread. 

We can evaluate long-term Emulator performance in more detail by examining the PDF of system state as a function of more than one variable. In addition to $U(\text{30 km})$ as a reduced variable indicative of vortex strength, \citeA{Finkel21} used the Integrated Heat Flux, IHF, as a second reduced variable indicative of wave forcing that threatens to destablize the vortex:
\[
\text{IHF}(z) 
=\int_{0}^{z}e^{-z'/H}\overline{v'T'}\,dz'
\propto\int_0^z|\Psi(z')|^2\frac{\partial\varphi}{\partial z'}\,dz',
\]
where we have decomposed the streamfunction into a magnitude and a phase, $\Psi(z)=|\Psi(z)|e^{i\varphi(z)}$. Following \citeA{Finkel21}, we will use IHF(30 km) as an effective variable for displaying and interpreting results. It captures geometric information about the streamfunction, namely its magnitude-weighted phase tilt between 0 and 30 km, as well as energetic information, namely the wave-induced northward thermal energy transport, upward wave activity transport, and downward momentum transport that weakens the vortex and triggers an SSW. Since the vertical distribution of zonal wind, $U(z)$, is predicted well by $U(\text{30 km})$, the variable pair ($U(\text{30 km})$, IHF(30 km)) is an effective dimensional reduction of the full 75-dimensional state of the Holton-Mass model.  

\begin{figure}[h!]
    \centering
    \includegraphics[width=1\linewidth]{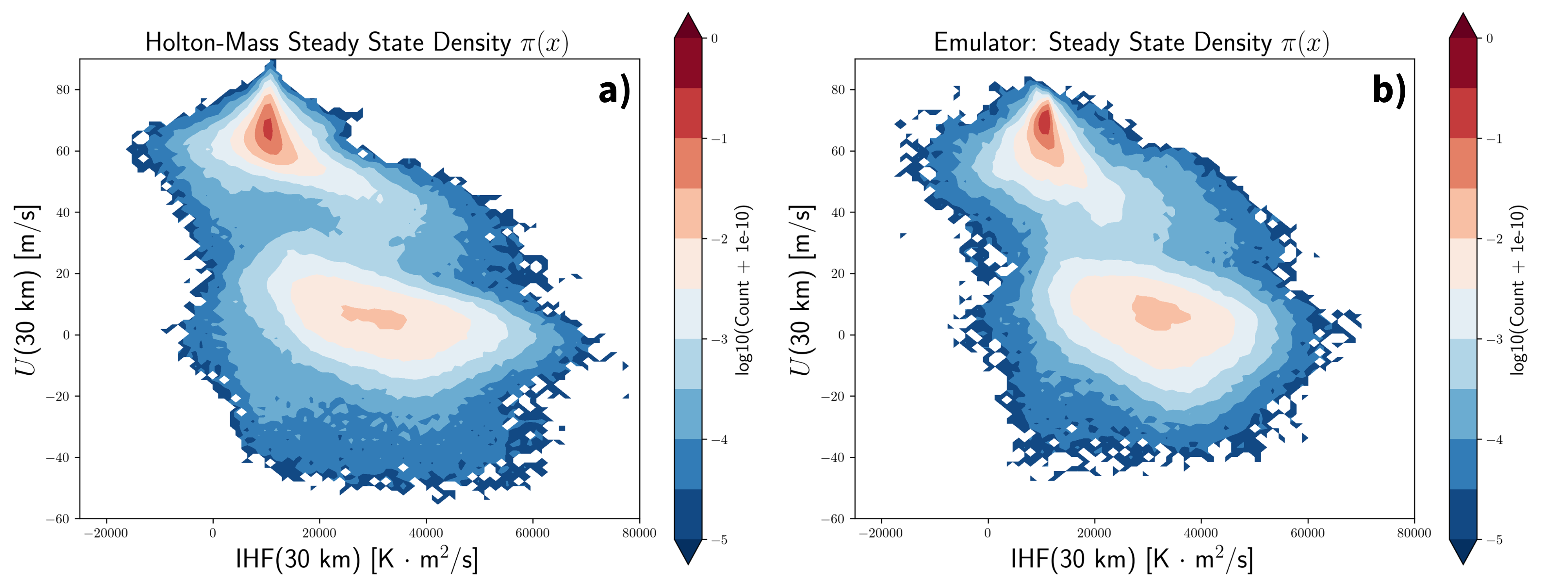}
    \caption{Two-dimensional projections of the PDF of the Holton-mass (left) and the Emulator (right) with respect to $U$(30 km) and IHF(30 km).}
    \label{fig:steadystate}
    \vspace{-1em}
\end{figure}

Figure \ref{fig:steadystate} shows the two-dimensional projection of the PDF of the Holton-Mass and the Emulator with respect to $U$(30 km) and IHF(30 km). The emulator slightly overestimates the strength of the strong vortex regime, placing the mode within $A$ a bit too high along the $U$(30 km) axis, but the additional context provided by plotting the density as a joint function of IHF(30 km) as well as $U$(30 km) increases our confidence in the emulator. The locations of both states in ($U(\text{30 km})$, IHF(30 km)) are well-captured by the Emulator, as are the shapes and orientations of the PDF contour lines. The Emulator does not reproduce the two-dimensional PDF of the Holton-Mass model perfectly, but this evaluation metric is an extremely high bar for emulator performance.

\vspace{-1em}

\newpage
In addition to projections of the PDF of the state of the system, it is important to examine the Emulator's representation of the system's long-term temporal evolution. We start by considering the \emph{regime persistence times} and \emph{transition durations}, defined as follows. A long timeseries $\mathbf{X}(t)$ will execute a sequence of transitions from $A$ to $B$ and back again, marked by a sequence of special timestamps 
\[
{\gamma_{\clab}^{(1)},\tau_{\clb}^{(1)},\gamma_{\clba}^{(1)},\tau_{\cla}^{(2)},\gamma_{\clab}^{(2)},\hdots}
\]
where transition paths start and end. Recalling $C=(A\cup B)^c$ is the region of intermediate vortex strength,
\begin{align}
    \mathbf{X}(t)&\in A && t=\gamma_{\clab}^{(1)} \notag\\
    \mathbf{X}(t)&\in C && \gamma_{\clab}^{(1)}<t<\tau_{\clb}^{(1)} \notag\\ 
    \mathbf{X}(t)&\in B && t=\tau_{\clb}^{(1)} \notag\\
    \mathbf{X}(t)&\in B\cup C && \tau_{\clb}^{(1)}\leq t\leq \gamma_{\clba}^{(2)} \notag\\
    \mathbf{X}(t)&\in B && t=\gamma_{\clba}^{(2)} \notag\\
    \mathbf{X}(t)&\in C && \gamma_{\clba}^{(2)}<t<\tau_{\cla}^{(2)}\notag\\
    \mathbf{X}(t)&\in A && t=\tau_{\cla}^{(2)} \notag\\ 
    \mathbf{X}(t)&\in A\cup C && \tau_{\cla}^{(2)}\leq t\leq \gamma_{\clab}^{(2)} \notag\\
    \mathbf{X}(t)&\in A && t=\gamma_{\clab}^{(2)} \notag\\
    \hdots& \notag
\end{align}
The $i$th \emph{regime persistence time} in $A$ is $\gamma_{\clab}^{(i)}-\tau_{\cla}^{(i)}$, and likewise for $B$. The $i$th \emph{transition duration} from $A$ to $B$ is $\tau_{\clb}^{(i)}-\gamma_{\clab}^{(i)}$, and likewise from $B$ to $A$. Each is a random variable, and their expected values are the long-term averages over $i$.

\begin{figure}[h!]
    \vspace{-2.5em}
    \centering
    \includegraphics[width=0.9\linewidth]{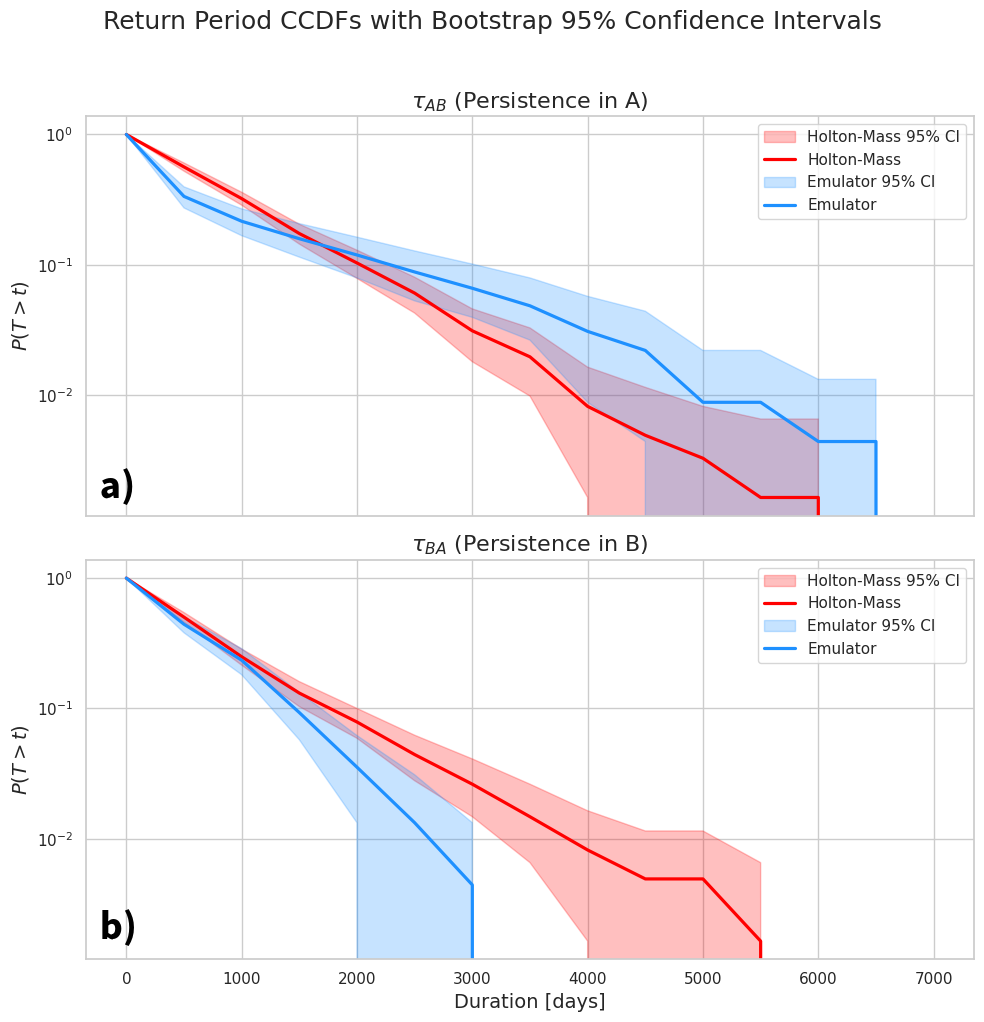}
    \vspace{-1em}
\caption{Complementary cumulative distribution functions (CCDFs) of persistence times for the Holton–Mass model (red) and the emulator (blue). Top: $\tau_{\clab}$, persistence in set A before transitioning to set B. Bottom: $\tau_{\clba}$, persistence in state B before transitioning to state A. Solid lines show the median empirical CCDFs, and the shaded regions indicate 95\% bootstrap confidence intervals based on 1000 resamplings. The figure uses fixed 500-day bins and logarithmic axes to highlight differences in the tail behavior of the persistence-time distribution.}
    \label{fig:returnperiods}
\end{figure}

Complementary cumulative distribution functions (Fig.~\ref{fig:returnperiods}) show that the Emulator accurately reproduces the structure of the exponential distribution of $\tau_{\clab}$ and $\tau_{\clba}$ exhibited by the Holton-Mass model into the tail of long persistence times. An important difference is that the Holton-Mass model shows similar persistence durations in sets $A$ and $B$, but the emulator shows slightly longer persistence durations in set $A$ than set $B$. A particular issue is that the emulator under-represents long-tail persistence in set $B$.

\begin{figure}[h!]
    \centering
    \includegraphics[width=1\linewidth]{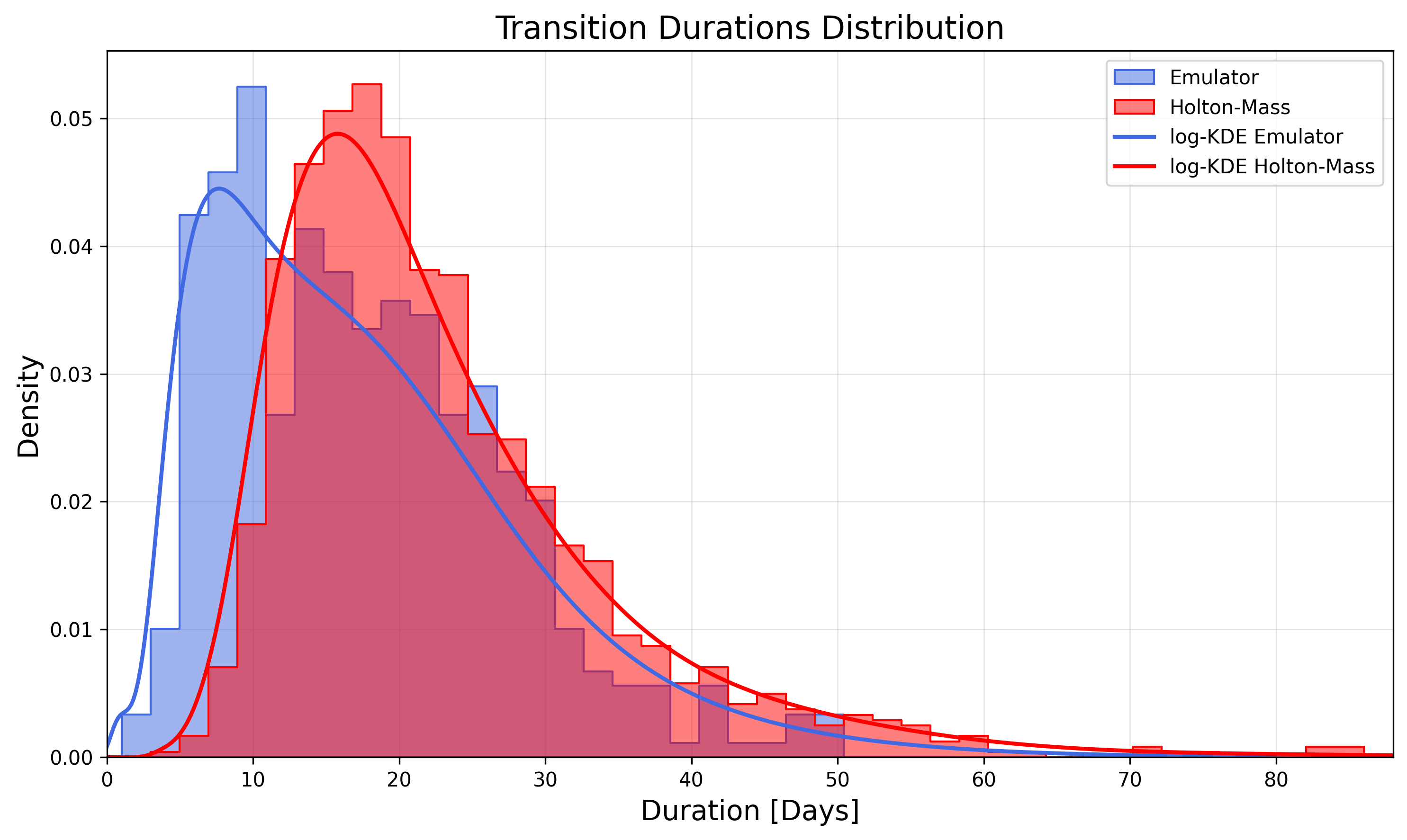}
    \caption{Distribution of transition durations between vortex regimes for the emulator (blue) and the Holton--Mass model (red). Bars are normalized histograms on shared bins. Solid curves are kernel density estimates (KDE) fit in log-duration space and mapped back to days as $P(t)=Q(\log t)/t$, where $Q$ is a kernel density estimate of the log durations; this construction guarantees the estimated density is zero for non-positive durations. The emulator's transition durations are skewed slightly toward faster transitions than the Holton--Mass model's.}
    \label{fig:transitiondurations}
\end{figure}

We also consider transition durations, which represent the ``suddenness'' of sudden stratospheric warming events and hence the typical time available to prepare once a transition has initiated. An accurate distribution of transition durations is therefore a key desideratum for the Emulator. When comparing the Emulator with the Holton-Mass model on this metric, we find that they have very similar distributions, with the emulator's distribution 
skewed slightly toward faster transition times (Fig.~\ref{fig:transitiondurations}), further supporting the effectiveness of the Emulator on long-term statistics.

\subsection{Committor and expected lead time}

In this section, we evaluate the emulator using two risk quantifiers: the committor and the expected lead time. The SSW committor, $q^+(x)$ is the probability that a partially weakened vortex (initially in state $x$) collapses before it recovers (reaches $B$ before $A$):
\begin{align}
    q^+(x)=\mathbb{P}\{\tau_{\clb}(t)<\tau_{\cla}(t)|\mathbf{X}(t)=x\},
\end{align}
where the first-hitting times $\tau_{\cla}$ and $\tau_{\clb}$ are as defined in section \ref{sec:data}. 
The expected time it takes the system to reach state $B$, \emph{given that it reaches $B$ first}, is called the expected lead time: 
\begin{align}
    \eta_B^+(x)=\mathbb{E}\big[\tau_{\clb}(t)|\tau_{\clb}(t)<\tau_{\cla}(t)\text{ and }\mathbf{X}(t)=x\big].
\end{align}
Because the system is autonomous, neither $q^+$ nor $\eta_B^+$ depends on $t$.  
We will evaluate these risk quantifiers as a function of the same two variables as the PDF: $U(\text{30 km})$ and IHF(30 km).

We calculate the SSW committor, $q^+$, from a $10^6$ day timeseries. By definition, $q^+=0$ in state $A$ and $q^+=1$ in state $B$. For positions in phase space not in either state $A$ or state $B$ we calculate the committor empirically by binning with respect to $U(\text{30 km})$ and IHF(30 km) and computing the fraction of trajectories originating in each bin that ultimately hit state $B$ before they hit state $A$.  

\begin{figure}[h!]
    \centering
    \includegraphics[width=1\linewidth]{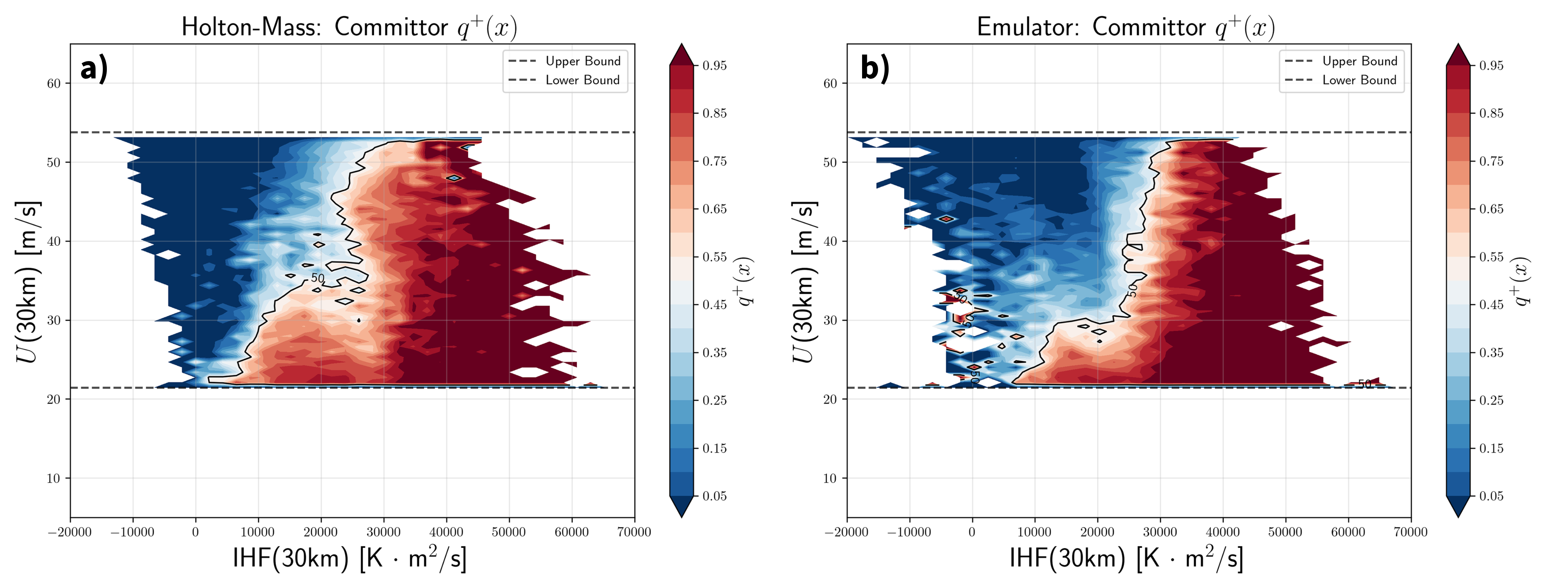}
    \caption{Committor functions calculated for the HM model (left) and the emulator (right) as a function of $U$(30 km) and IHF(30 km). The two dashed lines are the bounds of set $A$ and set $B$, respectively, and the black line is the level set $q^+=0.5$}
    \label{fig:committor}
\end{figure}

The emulator accurately reproduces the main features of the Holton-Mass model committor (Fig.~\ref{fig:committor}). In particular, it captures the joint dependence of the committor on $U(\text{30 km})$ and IHF(30 km): both smaller values of $U(\text{30 km})$ and larger values of IHF(30 km) contribute to an increased chance of hitting state $B$ before state $A$. This is interesting given that the states are defined by $U(\text{30 km})$ without reference to IHF(30 km), and it is notable that the emulator can reproduce this behavior. Physically, an increase in IHF(30 km) signifies a flow of thermal energy across space, which is a pre-requisite for the subsequent collapse in kinetic energy of the polar vortex  \cite{Yoden1987dynamical,Finkel2022data}

We calculate the expected lead time, $\eta^+_B$, using the same $10^6$ day timeseries as we used to calculate the committor, as well as the same bins in $U(\text{30 km})$ and IHF(30 km). In each bin we calculate $\eta^+_B$ is as the mean time from leaving that bin until reaching state $B$ for trajectories transitioning from state $A$ to state $B$ that pass through that bin. 

\begin{figure}[h!]
    \centering
    \includegraphics[width=1\linewidth]{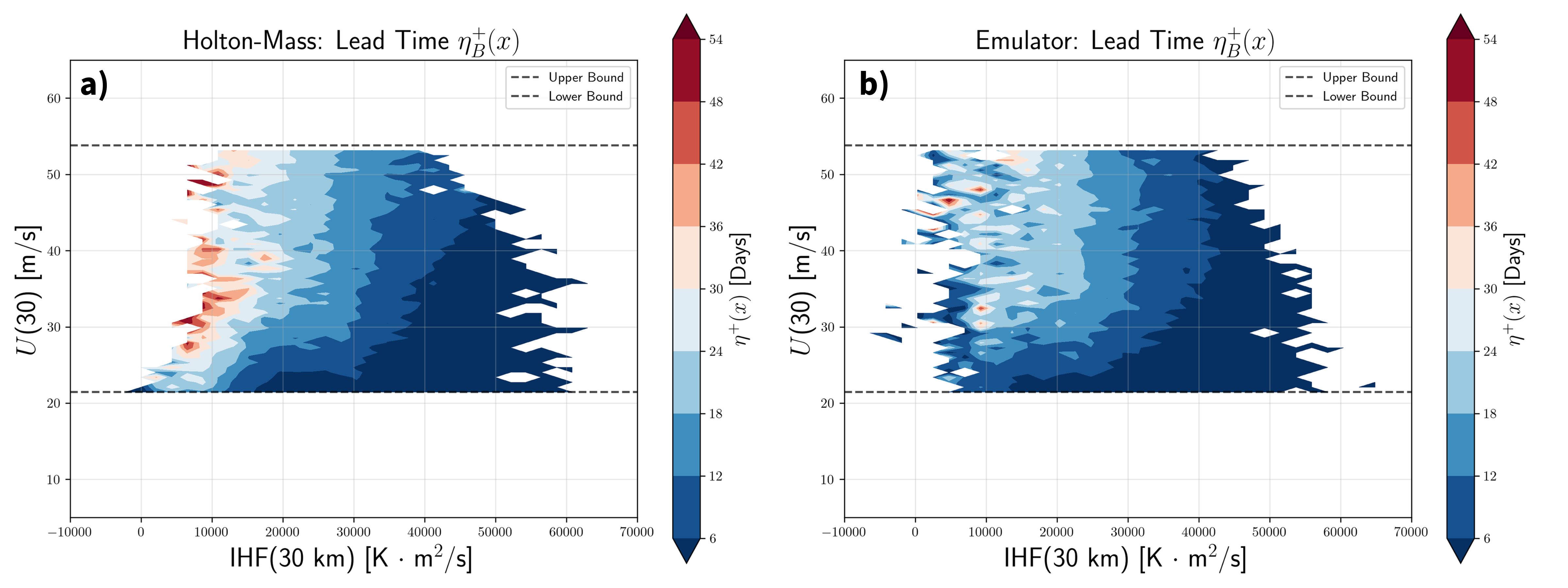}
    \caption{Expected lead time with respect to $U$(30 km) and IHF(30 km) of (left) the Holton-Mass Model, and (right) the emulator. The two dashed lines represent the bounds of state $A$ and state $B$, respectively. }
    \label{fig:leadtime}
\end{figure}

The emulator reproduces the basic trend shown by the Holton-Mass model that larger values of the committor function tend to be associated with shorter lead times, $\eta^+_B$, such that the lead time decreases as $U(\text{30 km})$ decreases and IHF(30 km) increases (Fig.~\ref{fig:leadtime}). The agreement of the emulator with the Holton-Mass model is striking for IHF(30 km) $>$ 2000 K$\cdot$m$^2\cdot$s$^{-1}$. For smaller values of IHF(30 km) the emulator underestimates the lead time, although our estimates of $\eta^+_B$ are noisy in this regime due to insufficient data. 

\vspace{-1em}

\subsection{Visualizing and Interpreting the Latent Space}
\label{sec:latent-space}

Thus far we have demonstrated that the emulator successfully captures the Holton-Mass model's metastable regime structure. We now interrogate the latent space to try to understand the way it encodes and distinguishes these four important classes. Recall that set $A$ is the strong vortex regime defined by zonal wind at 30 km greater than 53.8 m s$^{-1}$  (Eq.~\ref{eq:basin_A}) and set $B$ is the weak vortex regime defined by zonal wind at 30 km less than 21.4 m s$^{-1}$  (Eq.~\ref{eq:basin_B}). The strong-vortex class $\cla=\{t:\mathbf{X}(t)\in A\}$  can be sub-categorized as either
\begin{itemize}
    \item 
    transient: $\clab=\{t:\mathbf{X}(t)\in A,\mathbf{X}(t+1)\notin A,\text{ and }\tau_{\clb}(t+1)
    <\tau_{\cla}(t+1)\}$, meaning the  system is about to depart $A$ en route to $B$; or
    \item 
    persistent: $\claa=\{t:\mathbf{X}(t)\in A\text{ and }\tau_{\cla}(t+1)<\tau_{\clb}(t+1)\}$, meaning the system will either stay in $A$ or return before hitting $B$ next.
\end{itemize}
The categories $\clbb$ and $\clba$ are defined analogously, reversing the roles of $A$ and $B$.

\begin{figure}[H]
    \centering
    \includegraphics[width=\textwidth]{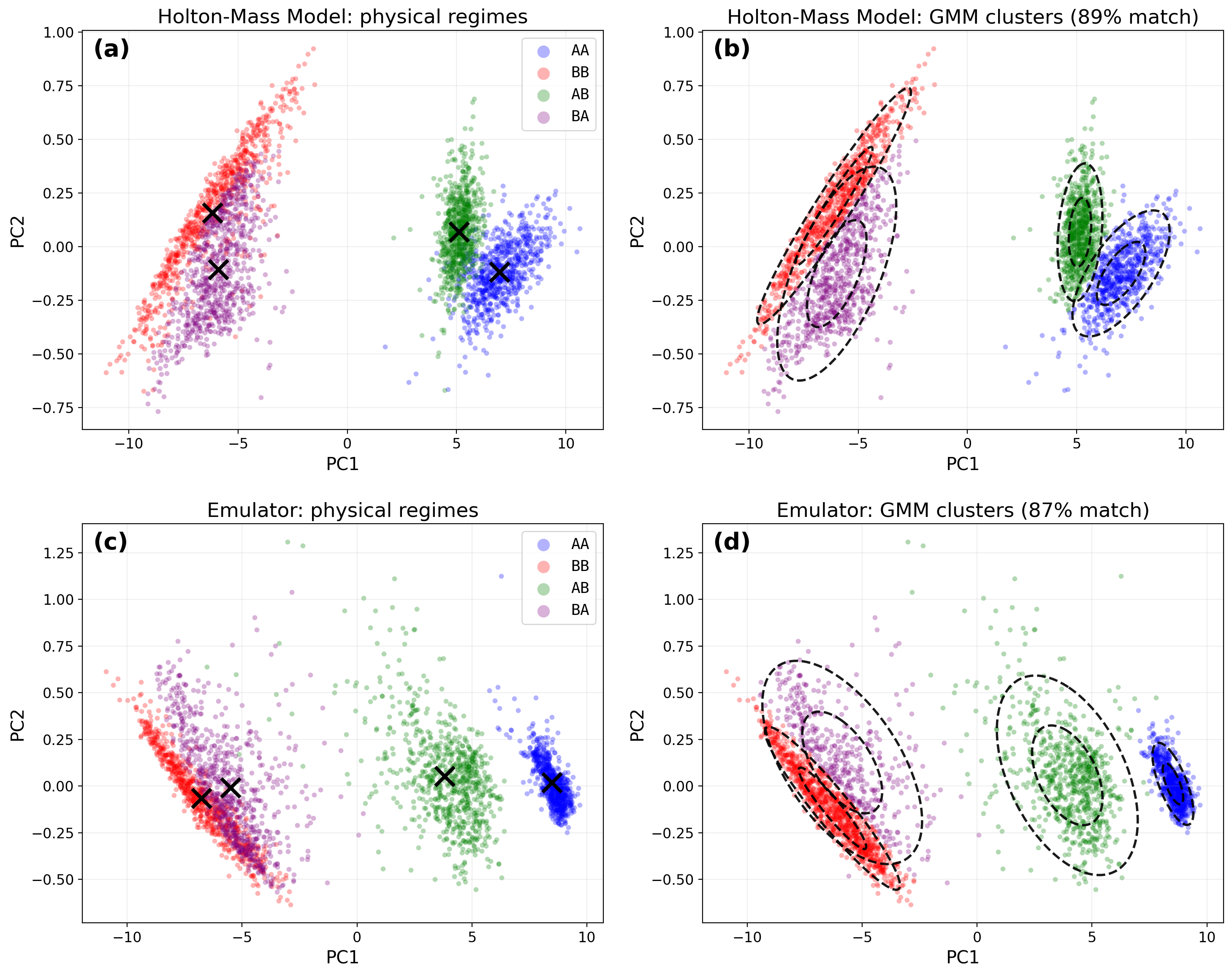}
    \caption{
    The four dynamical regimes emerge from the latent space without supervision. Latent mean vectors are projected onto the first two principal components for the Holton--Mass model (top) and the emulator (bottom). \textbf{Left:} points colored by their current and future physical regime classes---$\claa,\clab,\clbb$, and $\clba$---with $\times$ marking the regime centroids. \textbf{Right:} the same points colored by the cluster assigned by a four-component Gaussian mixture model fit to the latent coordinates alone (dashed curves show the $1\sigma$ and $2\sigma$ covariance ellipses), with each cluster colored by the physical regime it best matches. The title is the fraction of states whose cluster matches their regime. The unsupervised clusters closely reproduce the physically defined regime classes.}
    \label{fig:pca}
\end{figure}

From the $10^6$-day timeseries of the emulator and Holton-Mass model we sample
750 instances in each of the four classes discussed above. We perform this
analysis separately for each model: for the Holton-Mass model and the emulator
independently, we encode each of these 3000 input states, $x_{t}$, as 32-dimensional
latent mean and variance vectors using the CVAE, and fit a separate PCA to the latent
mean vectors of each model. We obtain similar results with the latent
variance vectors. Because each PCA is fit independently, the principal axes (and hence
the PC1/PC2 coordinates shown in Fig.~\ref{fig:pca}) are defined per-model and are not directly comparable between the two. The first three principal components explain approximately
$99.86\%$ of the total variance in the Holton-Mass Model (PC1 $\approx 99.60\%$,
PC2 $\approx 0.17\%$, PC3 $\approx 0.09\%$) and $99.87\%$ in the emulator
(PC1 $\approx 99.65\%$, PC2 $\approx 0.14\%$, PC3 $\approx 0.09\%$).

The leading principal component, PC1, separates class $\cla=\claa\cup\clab$ well from class $\clb=\clbb\cup\clba$ (Fig.~\ref{fig:pca}).  Including PC2 or PC3 is necessary for making finer distinctions within a regime. Classes $\clbb$ and $\clba$ occupy the same range of PC1
and are distinguished only by the combination of PC1 and PC2 or PC3. For example, PC1 and PC2 separate the four classes into fairly distinct clusters (Fig.~\ref{fig:pca}). Interestingly, the CVAE is just as effective at performing this classification for the Holton-Mass model as for the emulator. The (PC1,PC3) combination behaves similarly (not shown). This suggests that the encoder uses a secondary principal component to determine whether the system, starting in set $A$ or $B$, is likely to remain there for some time or imminently go through a transition. The secondary separation between persistent and transient classes is not perfect, which may be due to information contained in additional principal components. But even if a moderate number of PCs is needed, this result is a promising potential demonstration of a \emph{method for identifying} precursors of SSWs and other regime transitions. We stress that the specific precursor discovered here only applies to the idealized Holton-Mass model, so will not carry over to real-world SSWs. What may carry over is the method of (1) building a stochastic emulator from real-world data or GCM output, (2) dimensionality reduction and clustering in the latent space, and (3) matching the clusters to physically meaningful regimes. If the method holds up for real-world SSWs at the subseasonal-to-seasonal timescale, it could be used to build an early warning system for predicting these high-impact events. It would of course be necessary to interrogate the discovered clusters rigorously for physical interpretation and causal significance in order to build confidence in the result.

Crucially, this latent space interpretation does not require class labeling. Fitting
a Gaussian mixture model (GMM) \cite{dempster1977em} to the latent representation, using only
the latent coordinates and withholding the class labels, recovers the same four groups
(Figure~\ref{fig:pca}, right). A GMM groups the data into a chosen number of clusters, each
described by a Gaussian cluster that can have its own size and orientation, and assigns every state to the cluster it most likely belongs to (whichever Gaussian density is largest at that point in latent space). The number of clusters is not imposed. A
standard label-free criterion, the silhouette score  \cite{rousseeuw1987silhouettes},  is independently maximized at $k=4$
for both the Holton--Mass model and the emulator. When only PC1 is supplied, the same criterion instead selects two clusters,
corresponding to classes $\cla$ and $\clb$, and cannot discern the persistent from the transient classes ($\claa$ vs. $\clab$ and $\clbb$ vs. $\clba$). This information resides in the higher principal
components, but mostly just in the first two: in the PC1-PC2 plane shown, $\approx 87$--$89\%$ of states fall in the cluster matching their
physically defined regime, improving to $\approx 96\%$ (emulator) and $\approx 89\%$
(Holton--Mass) when the full $32$-dimensional latent space is used, distinguishing classes $\claa$ from $\clab$ at $98$--$100\%$ purity and only somewhat mixing together the weak-vortex classes $\clbb$ and $\clba$. Therefore, it is possible to identify the
four dynamical regimes from the latent space alone and, mapping them back to physical
space, recover the strong- and weak-vortex classes and their persistent vs. transient sub-classes without
supervision. The mixture model also returns the posterior probability that each state belongs
to its assigned cluster. These assignments are nearly deterministic for classes $\cla$ and the
$\clab$ (mean maximum posterior probability of $0.99$, with
$98$--$100\%$ of states assigned above $0.9$) and somewhat softer for classes $\clbb$ and
$\clba$ (median probabilities of $0.92$ and $0.95$), so the residual
assignment uncertainty is concentrated at the $\clbb/\clba$ boundary.

\begin{figure}[h!]
    \centering
    \includegraphics[width=\textwidth]{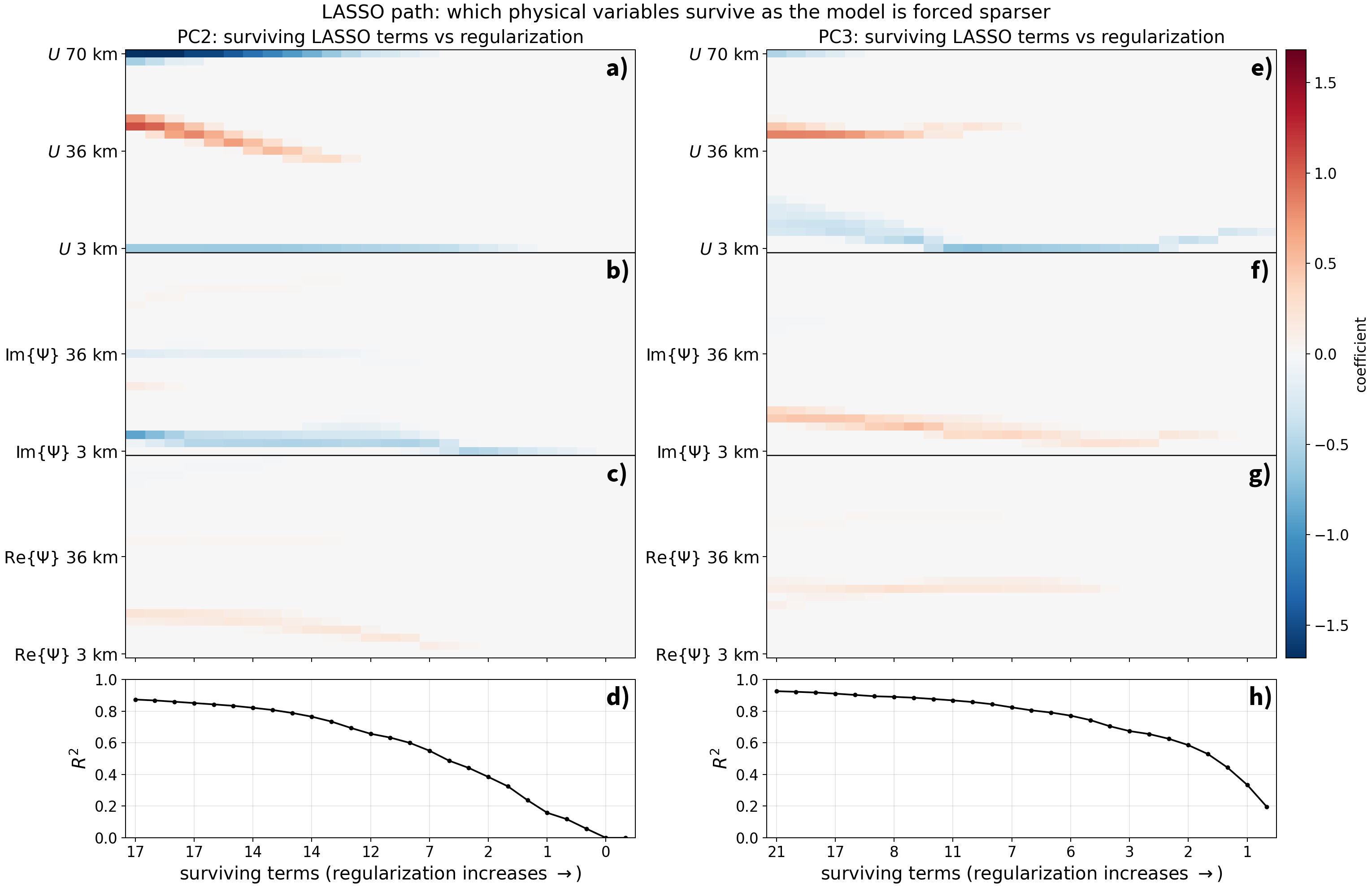}
    \caption{LASSO regularization paths for PC2 (left) and PC3 (right) of the emulator's
    latent space. Each row is one of the 75 state variables, grouped by field (real and
    imaginary parts of the streamfunction $\Psi$ and the zonal wind $U$) and ordered by
    altitude; columns run from weak to strong regularization, labeled by the number of
    surviving (nonzero) coefficients; color is the standardized regression coefficient.
    A variable stays colored as long as the model retains it, so the rows still colored
    toward the right are those kept under strong regularization. For both PCs the
    variables that survive the strongest regularization are at low altitude; the prominent
    upper-level zonal-wind bands in the least-regularized fits are dropped first,
    indicating that pattern is spread across many correlated levels rather than carried
    by a single essential variable.}
    \label{fig:lasso_path}
\end{figure}

PC1 is relatively straight-foward to understand physically. Across a random sample of emulator states, PC1 and $U(30\,\mathrm{km})$ are correlated at $r = 0.98$. Given that $U(30\,\mathrm{km})$ is the variable that defines regimes $A$ and $B$, this lends confidence in our latent space analysis. In order to better understand PC2 and PC3, we regressed these PCs from the emulator onto the 75 components of the state vector using LASSO \cite{tibshirani1996lasso}, which
fits a linear model while regularizing to drive as many coefficients as possible to zero (Fig.~\ref{fig:lasso_path}). Both PC1 and PC2 are largely linear functions of the state ($R^2 = 0.91$ and $0.94$ for PC2
and PC3, respectively), with most of the weight on the zonal wind profile. PC3 admits a
compact description: roughly ten terms, combining the zonal wind at low and middle
levels with the streamfunction in the lower stratosphere, retain $R^2 \approx 0.89$.
PC2 instead draws on the zonal wind across many vertical levels and does not reduce to
a small number of terms as well~(Fig.~\ref{fig:lasso_path}).  Moreover,  PC3 is moderately correlated with IHF(30 km)
($r = 0.66$, reduced to $r = 0.42$ after removing the linear dependence of both
variables on $U(30\,\mathrm{km})$), whereas PC2 is only weakly correlated with IHF.
Taken together, these results suggest the following physical interpretation of the latent space: PC1
represents the vortex strength (essentially the zonal wind at 30 km), PC3 represents the
wave forcing that drives transitions (the wave amplitude and heat flux that build
before the vortex breaks down), and PC2 (more speculatively) may represent the wind profile shape.

\section{Discussion}

The PCA of the latent space shows that our CVAE emulator does more than match short-term forecasts and long-term statistics: it learns a low-dimensional, dynamically meaningful representation of the Holton–Mass system. By projecting the 32-dimensional latent means onto the leading principal components, we find four well-separated clusters associated with strong and weak vortex configurations and the onset of transitions, effectively revealing an emergent potential early warning signal for SSW-like regime changes. Our results demonstrate that VAEs can encode slow dynamical variables and fast transition pathways in their latent space in a stochastic, metastable model where rare transitions are central to the problem. 
This provides a concrete, data-driven handle on the geometry of regime transitions, and for the field of interpretable ML, it offers an example where a relatively simple probabilistic emulator yields a latent manifold that maps cleanly onto physically defined regimes rather than remaining a black box.

This emergent organization indicates that the ResNet-CVAE captures not only the two stable states of the HM model but also the dynamical pathways connecting them. The fact that such structure arises without regime labels during training suggests that by learning the conditional distribution $p(\mathbf{X}(t+1)\!\mid\!\mathbf{X}(t))$, the network implicitly discovers physically meaningful manifolds of variability. This supports the broader motivation of this work: well-designed generative emulators can replicate stochastic dynamics while also revealing internal representations that mirror the underlying physical regime structure.

In a broader context, this work fits alongside emerging efforts to use AI both to accelerate climate simulations and to better understand extremes. Recent reviews emphasize that AI-based methods for climate extremes will only be trusted if their speed is combined with physical fidelity and interpretability \cite{Materia2024,lai2025machine}. Large-scale climate emulators such as ACE2 and LUCIE \cite{WattMeyer2025,guan2025lucie} show that autoregressive ML models can stably emulate complex atmospheric dynamics, though their internal representations are often opaque. At the same time, generative AI approaches such as FM-Cast \cite{Tao2026efficient} for SSW ensemble prediction demonstrate how probabilistic deep learning can rival or exceed operational systems for real-world events while remaining computationally efficient. 

Our emulator contributes at the ``idealized building-block'' level: it demonstrates, in a controlled setting provided by the Holton–Mass model, that a carefully designed stochastic deep learning model can faithfully capture transition statistics and SSW risk quantifiers while exposing a latent space in which regimes and precursors are clearly organized. Researchers working on operational SSW prediction and AI weather models may be most interested in how this latent structure could inform feature design, regime-aware training, or rare-event sampling strategies in more realistic systems, while ML theorists and climate dynamicists may see it as a testbed for future work on disentangled or physics-informed latent variables that enable targeted manipulation of SSW likelihood or timing.

\section{Conclusion}

Here we have demonstrated that a ResNet-Inspired Conditional VAE is able to accurately emulate the stochastic metastable Holton-Mass model of sudden stratospheric warmings. Although we trained only on one-day predictions, the emulator is able to match both short-term and long-term statistics, including the two-dimensional structure of important variables such as the state PDF, transition committor function, and transition expected lead time. Moreover, a PCA analysis of the Emulator's latent space reveals four clusters associated with the two metastable states and the two transition pathways. This work serves as a promising demonstration of an effective analysis method to build an interpretable emulator of a stochastic climate model for subseasonal to seasonal forecasting and extreme event early warning.

 \section*{Open Research Section}
 The code to reproduce the work in this paper is archived in a public repository \cite<>[available at \url{https://doi.org/10.5281/zenodo.21144536}]{modified_code}.
\section*{Conflict of Interest declaration}
The authors declare there are no conflicts of interest for this manuscript.

\acknowledgments
This work was funded in part by the United States National Science Foundation through award NSF RISE–2425898. We acknowledge the support of undergraduate summer Polsky Research Fellowships from the Institute for Climate and Sustainable Growth at the University of Chicago. We acknowledge support from a University of Chicago Quad Undergraduate Research Scholarship. JF, PH, and DSA acknowledge funding from the Institute for Climate and Sustainable Growth at the University of Chicago. AC was supported by NSF 2425667. JF acknowledges support from DSI-AICE fellowship from the University of Chicago AI for Climate Initiative. Feedback from Ed Gerber and two anonymous reviewers helped us significantly improve this paper. 


%
%


%
%
%
%
%

\medskip

\bibliography{references}

\end{document}